\documentclass{optica-article}

\journal{opticajournal} 

\articletype{Research Article}

\usepackage{lineno}
\usepackage{siunitx}
\DeclareSIUnit\angstrom{\text {Å}}

\begin{document}

\title{Nanoscale Dark-Field Imaging in Full-Field Transmission X-Ray Microscopy}

\author{Sami Wirtensohn,\authormark{1,2,3,4*}, Peng Qi\authormark{5}, Christian David\authormark{5}, Julia Herzen\authormark{2, 3, 4}, Imke Greving\authormark{1}, Silja Flenner\authormark{1}}

\address{\authormark{1}Institute of Materials Physics, Helmholtz-Zentrum Hereon, Max-Planck-Straße 1, 21502 Geesthacht, Germany\\
\authormark{2}Research Group Biomedical Imaging Physics, Department of Physics, TUM School of Natural Sciences, Technical University of Munich, James-Franck-Straße 1, 85748 Garching, Germany \\
\authormark{3}Chair of Biomedical Physics, Department of Physics, TUM School of Natural Sciences, Technical University of Munich, James-Franck-Straße 1, 85748 Garching, Germany \\
\authormark{4}Munich Institute of Biomedical Engineering, Technical University of Munich, Boltzmannstraße 11, 85748 Garching, Germany\\
\authormark{5}Paul Scherrer Institut, Forschungsstrasse 111, 5232 Villigen, Switzerland}

\email{\authormark{*}sami.wirtensohn@hereon.de} 

\begin{abstract*}
The dark-field signal uncovers details beyond conventional X-ray attenuation contrast, which is especially valuable for material sciences. In particular dark-field techniques are able to reveal structures beyond the spatial resolution of a setup. However, its implementation is yet limited to the micrometer regime. Therefore, we propose a technique to extend full-field transmission X-ray microscopy by the dark-field signal. The proposed method is based on a well-defined illumination of a beam-shaping condenser, which allows to block the bright-field by motorized apertures in the back focal plane of the objective’s lens. This method offers a simple implementation and enables rapid modality changes while maintaining short scan times, making dark-field imaging widely available at the nanometer scale.
\end{abstract*}

\section{Introduction}
The hierarchical 3D structure and inhomogeneities of a sample are of particular interest in different scientific fields such as materials sciences, biomedical research, and geology. Hence, transmission X-ray microscopy (TXM) has become an invaluable tool for investigating sample properties down to the nanometer scale~\cite{Andrews2011, Spence2021, Longo2020, Flenner2020}. The TXM is a full-field imaging technique that provides structural information in real space based on the attenuation properties of the sample. For low attenuating samples, the contrast can be improved by utilizing the phase shift induced by the sample structures and composition. This can be implemented at a TXM setup by Zernike phase-contrast~\cite{Zernike1942, Schmahl1994, Neuhausler2003, Stampanoni2010, Flenner2023}. If smaller structures are of interest, a third imaging modality is needed, which is based on the scattering caused by the sample. Scattering information is usually gained by performing small-angle X-ray scattering (SAXS)~\cite{Pilz1979, Giuntini2021}, wide-angle X-ray scattering (WAXS)~\cite{Buffet2012, Graewert2013, Smith2021}, or other diffraction experiments. To obtain spatial information about the scattering structures, the sample has to be scanned while the step size determines the spatial resolution~\cite{Li2022}. This is a powerful but very time-consuming technique, often not applicable to tomography. An approach utilizing the scattering information for imaging is based on the magnification of single Bragg peaks \cite{Simons2015, Yildirim2020}. Here, however, the sample needs to be scanned over multiple different angles around two rotation axes, making the imaging process slow. This technique is also limited to (poly)crystalline samples.

A full-field imaging technique capable of visualizing scattering structures has emerged with the introduction of grating-based X-ray interferometry in the mirco-regime: X-ray dark-field imaging~\cite{Pfeiffer2008}. It has proven to be a versatile technique widely used for applications in medical research, such as lung studies \cite{Willer2021, Urban2022, Gassert2021} and breast studies \cite{Heck2020, Scherer2016, Auweter2014}, the validation of manufacturing processes \cite{Ozturk2023, Endrizzi2015, Revol2011} and the advancement of materials science analysis \cite{Blykers2022}. The dark-field signal is sensitive to electron density changes, in the range below the spatial resolution of the imaging system applied. Therefore, sample inhomogenities can be visualized, which would otherwise be undetectable by the X-ray attenuation contrast~\cite{Taphorn2023}. The additional modality complements the compositional and density information provided by conventional X-ray attenuation and phase contrast imaging. Its different contrast mechanism emphasizes certain sample features and thus increases detectability. This allows the study of the overall hierarchical structure of samples.

While grating-based implementations of the dark-field are especially prominent in micro-computed tomography (micro-CT), the nano regime presents a distinct challenge, since gratings cannot be manufactured with the desired pitch and height. The additional X-ray optics in a TXM also cause further challenges like the twin image problem, which is created by higher diffraction orders of the gratings~\cite{Takano2018a}. 
A different approach for the nanometer regime is presented in the theoretical work by S. Vogt \textit{et al.}~\cite{Vogt2001}. They simulated an approach to dark-field X-ray microscopy, inspired by setups analogous to optical light microscopy. They proposed adding an aperture ring in front of a Fresnel zone plate (FZP) as a collimator and a ring stop in the objective's back focal plane (BFP). However, this setup has not been shown to work experimentally.
The study by Yin \textit{et al.} \cite{Yin2006} considered implementing the above approach based on a capillary condenser, but faced significant challenges in aligning the capillary condenser with the ring stop due to positional fluctuations of the horizontal and vertical illumination. Therefore, Yin \textit{et al.} \cite{Yin2006} proposed a different approach to implement dark-field imaging in a TXM setup. They suggest using an FZP with a reduced numerical aperture (NA) to focus only the scattered light. The bright field passes by outside the FZP and is therefore not focused on to the detector. However, their approach still raises the concern of a reduced flux, potentially limiting the applicability of the technique. Furthermore, it does not allow to change between the attenuation and dark-field signal without modifying the setup.

To address the limitations of these existing approaches, we propose a new full-field dark-field TXM method. It takes advantage of the well-defined pattern in the BFP of the FZP created by a beam-shaping condenser~\cite{Jefimovs2008}. Since diffracted and undiffracted parts are spatially seperated in the BFP, the detector image can be modified by closing motorized apertures. This approach provides high flux and stability, extending the dark-field imaging capabilities into the nanometer regime.

\section{Method}
\subsection{Transmission X-ray Microscopy}
The experiments are performed at the P05 nanotomography end station at the PETRA III storage ring at DESY, Hamburg, operated by the Helmholtz-Zentrum Hereon.
The X-ray beam is generated by a \SI{2}{\meter} long U29 undulator with a source size of \SI{36.0}{\micro\meter} × \SI{6.1}{\micro\meter} and a divergence of \SI{28.0}{\micro\radian} × \SI{4.0}{\micro\radian}. The beam is monochromatised by a Si-111 double crystal monochromator at an energy of \SI{11}{\kilo\electronvolt}.
As a detector, a Hamamatsu C12849-101U with a pixel size of \SI{6.5}{\micro\meter}, a chip size of 2048 x 2048 pixels, and a \SI{10}{\micro\meter} Gadox scintillator is used. The images are read out with a depth of 16 bits \cite{Flenner2022p}. The detector is placed in the next experimental hutch, approximately \SI{19}{\meter} behind the sample.

The TXM setup is based on a beam-shaping condenser \cite{Jefimovs2008}. It consists of several fields of gratings with constant periods, arranged in a circle. These gratings diffract the incoming parallel beam and converge it in the sample plane. There, the light from all gratings overlaps and creates a square-shaped flat-top illumination with the size of \SI{100}{\micro\meter} $\times$ \SI{100}{\micro\meter}. For all experiments, an iridium-based beam-shaping condenser with a diameter of \SI{1.8}{\milli\meter} and a finest structure size of \SI{50}{\nano\meter} is used. In the center behind the beam-shaping condenser, a beamstop with the diameter of \SI{1.1}{\milli\meter} is placed to block the direct beam, creating a hollow illumination. An order sorting aperture (OSA) blocks the higher diffraction orders of the condenser. The objective lens is a gold FZP with a diameter of \SI{300}{\micro\meter} and an outer zone width of \SI{50}{\nano\meter}~\cite{Flenner2020}. The optics are designed and manufactured by the Paul Scherrer Institute. 
The sample is placed in front of the FZP by an object distance $g$ which can be determined by the thin lens equation. For long detector distances $b$, this distance can be approximated by
\begin{align}
    \frac{1}{g} = \frac{1}{f_{\text{FZP}}} - \frac{1}{b} \approx \frac{1}{f_{\text{FZP}}}.
\end{align}
The described setup has a field of view (FoV) of \SI{93.27}{\micro\meter} with an effective pixel size of \SI{45.54}{\nano\meter} and reaches a resolution of \SI{114.42}{\nano\meter} in the projections based on the 20\% criterion of the modulation transfer function of a Siemens star projection. The resolution in the reconstructed slices is \SI{116.84}{\nano\meter} and was determined by the Fourier ring correlation~\cite{VanHeel2005}.

\subsection{Zernike Phase-Contrast Transmission X-ray Microscopy}
For the proposed method, it is helpful to understand Zernike phase contrast imaging, in which the illumination is also modified in the BFP of the FZP.\\
The method is frequently used and can be easily implemented in a TXM. In this technique, a $\frac{\pi}{2}$-shifting phase ring is placed in the BFP. In the BFP, the objective lens creates an image of the beam-shaping condenser, which consists of focal spots lying on concentric rings (Figure \ref{fig:setup} B, only one ring displayed for clarity). Here, the part of the beam that is refracted by the sample is spatially separated from the background and will therefore not pass through the phase-shifting phase rings. The phase-shifted and unshifted parts of the beam interfere in the detector plane and create a contrast-enhanced image~\cite{Vartiainen2014}.

\subsection{Dark-Field Transmission X-ray Microscopy}
\begin{figure}[htbp]
    \centering\includegraphics[width = \textwidth]{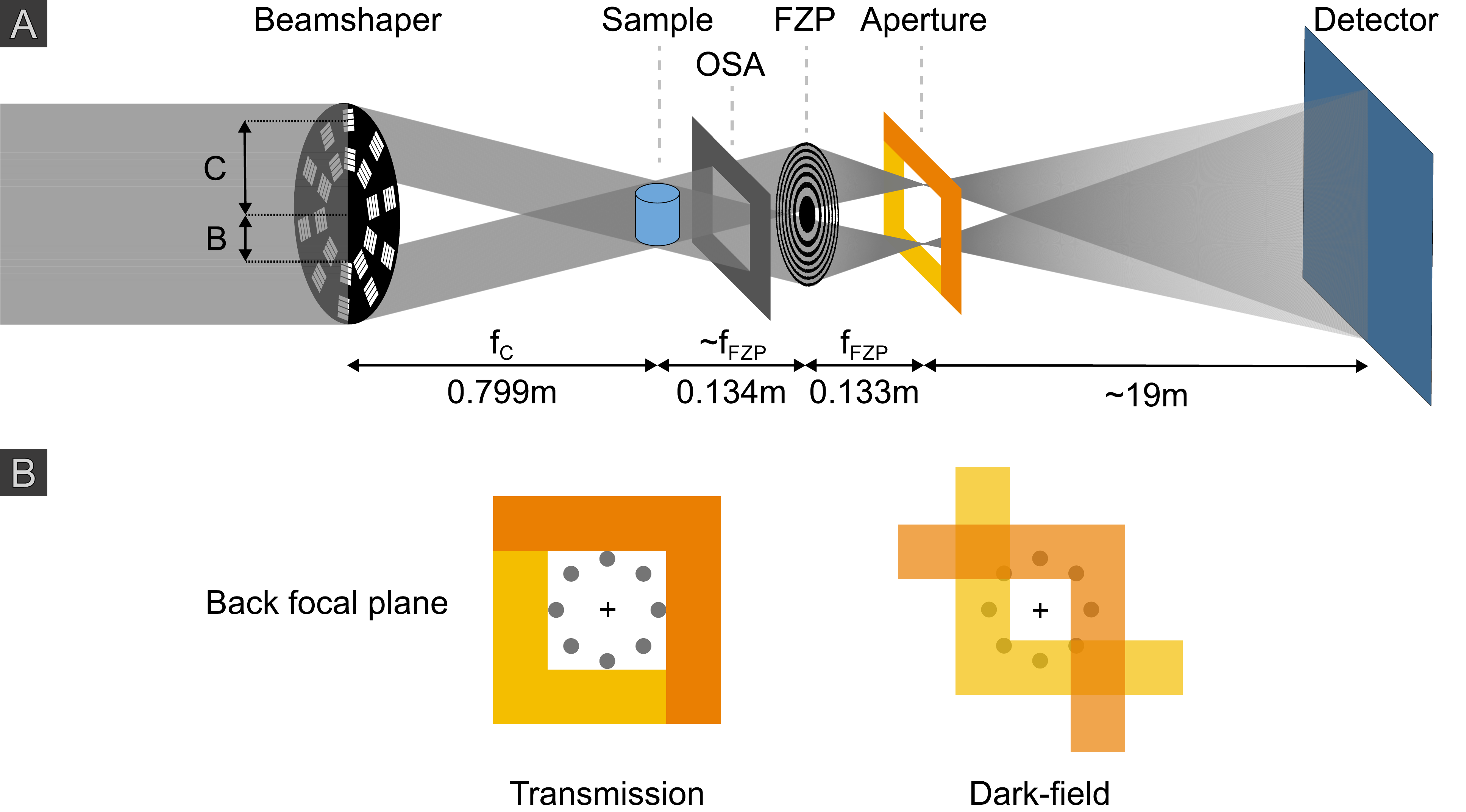}
    \caption{\textbf{A}: Schematic of the TXM dark-field setup. The beam-shaping condenser splits the beam into multiple deflected parallel beams, creating a ring of focused points in the back focal plane of the Fresnel zone plate (shown in A with only two beams). \textbf{B}: The two L-shaped apertures (yellow and orange) can be closed until all focal spots (grey dots) are fully blocked, switching from the transmission to the dark-field image.}
    \label{fig:setup}
\end{figure}

In order to obtain the dark-field signal in a TXM setup and thus information about the scattering structures of a sample, the bright field, i.e. the non-scattered part of the intensity, must be separated from the scattered intensity. The simplest method is to use the knowledge about the illumination, analogous to Zernike phase contrast, to block the bright field. 
The method proposed here, shown in Figure \ref{fig:setup} A, uses a set of motorized L-shaped apertures (marked yellow and orange). They are placed in the BFP and can be closed until all focal spots are obscured (Figure \ref{fig:setup} B). This fully blocks the bright field and only allows the scattered intensity to pass to the detector, producing the dark-field image. The apertures can also be opened, covering none of the focal spots and letting the whole light pass to the detector. In this mode, the microscope shows the normal transmission image.

\subsubsection{Scattering Vector}
\begin{figure}[htbp]
    \centering\includegraphics[width = \textwidth]{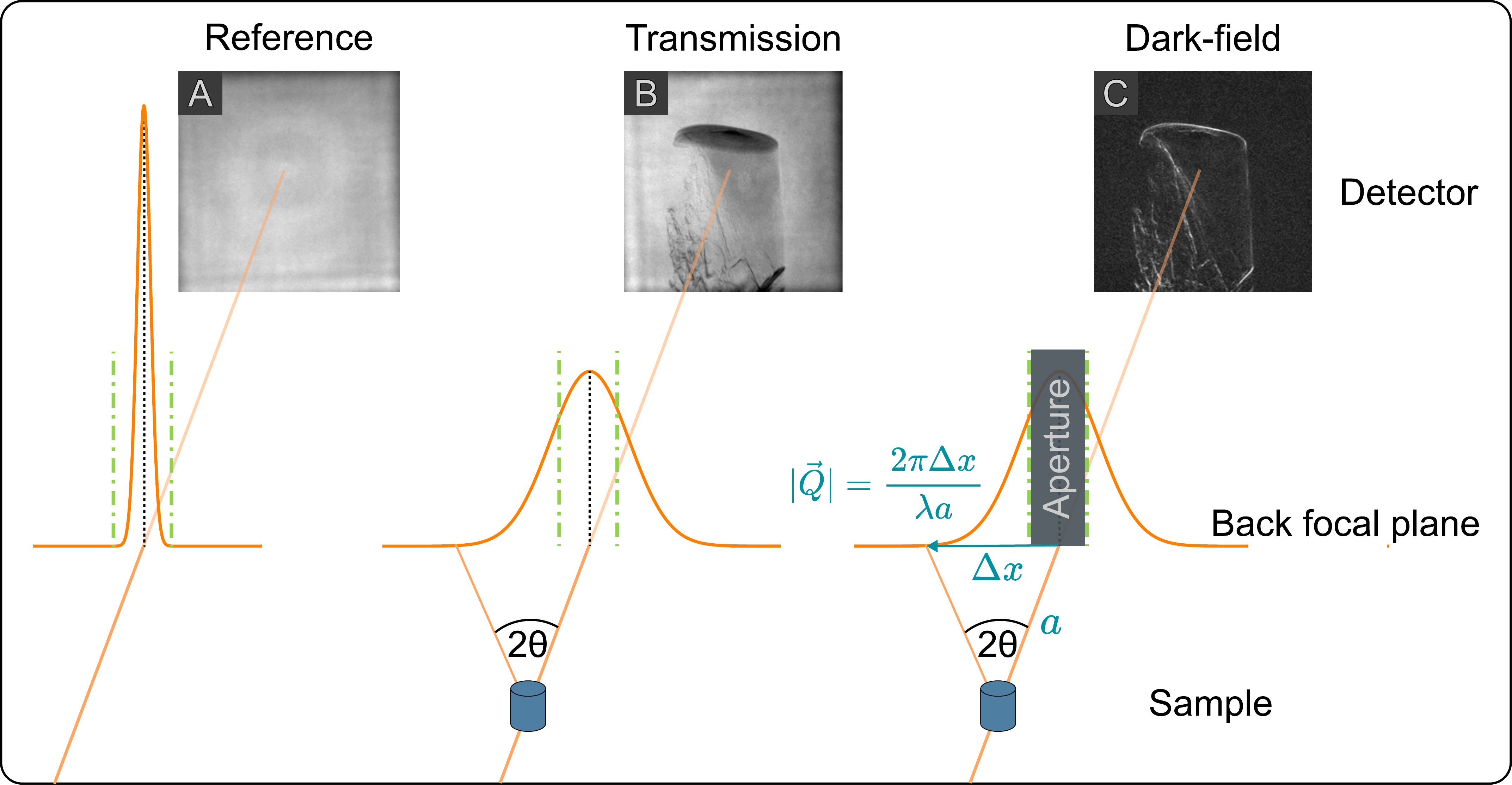}
    \caption{Intensity distribution of a focal spot in the back focal plane of the Fresnel zone plate. On the left is a focal spot without a sample in the beam (\textbf{A}). The spot has a defined peak position (black dotted line) and a defined total width (green dotted line). By introducing a scattering sample into the beam, the total width of the focal spot increases as shown in the center (\textbf{B}). By blocking the light at the position of the reference focal spot, only the scattered light can reach the detector, producing the dark-field image shown on the right (\textbf{C}).}
    \label{fig:gaussian_peak} 
\end{figure}

To better understand the formation of the TXM dark-field signal, consider one focal spot in the BFP of the FZP created by a single grating of the condenser with a Gaussian intensity distribution. It has a specific width and height, as shown in Figure \ref{fig:gaussian_peak} A. Introducing a sample into the beam reduces the total intensity due to attenuation and increases the width of the peak due to scattering in the sample (Figure \ref{fig:gaussian_peak} B). Since the illumination pattern without the sample is known, it can be blocked by an aperture. Hence only the scattered intensity can pass through, creating the dark-field image in the detector plane (Figure \ref{fig:gaussian_peak} C). In reality, the beam-shaping condenser produces a more complex illumination pattern with multiple circularly arranged focal spots, making it difficult to block each spot individually. Therefore, two L-shaped apertures are used, which are closed until all focal spots originating from the beam-shaping condenser are blocked. Although the outermost foci are relatively far from the aperture, they still contribute to the dark-field image by providing intensities with larger scattering angles, thus information about small sample features. The foci closer to the aperture contribute mainly intensities with small scattering angles, hence information from larger sample features. Due to the close relationship between the dark-field signal and scattering experiments, the contribution of a focal spot to the scattered intensity can be described by the magnitude of the scattering vector $|\Vec{Q}|$, which is given by

\begin{align}\label{eq:Q}
    |\Vec{Q}| = \frac{4\pi \sin{\theta}}{\lambda},
\end{align}

where $2\theta$ is the scattering angle \cite{Singh2017, Gommes2021}.
In order to determine the properties of the setup, the maximum scattering vector $|\Vec{Q}_{\text{max}}|$ has to be specified. It provides an idea of the total $Q$-range accepted by the setup and allows to evaluate the size of the smallest detectable structures in the sample.

\begin{figure}[htbp]
    \centering
    \includegraphics[width = 0.6 \textwidth]{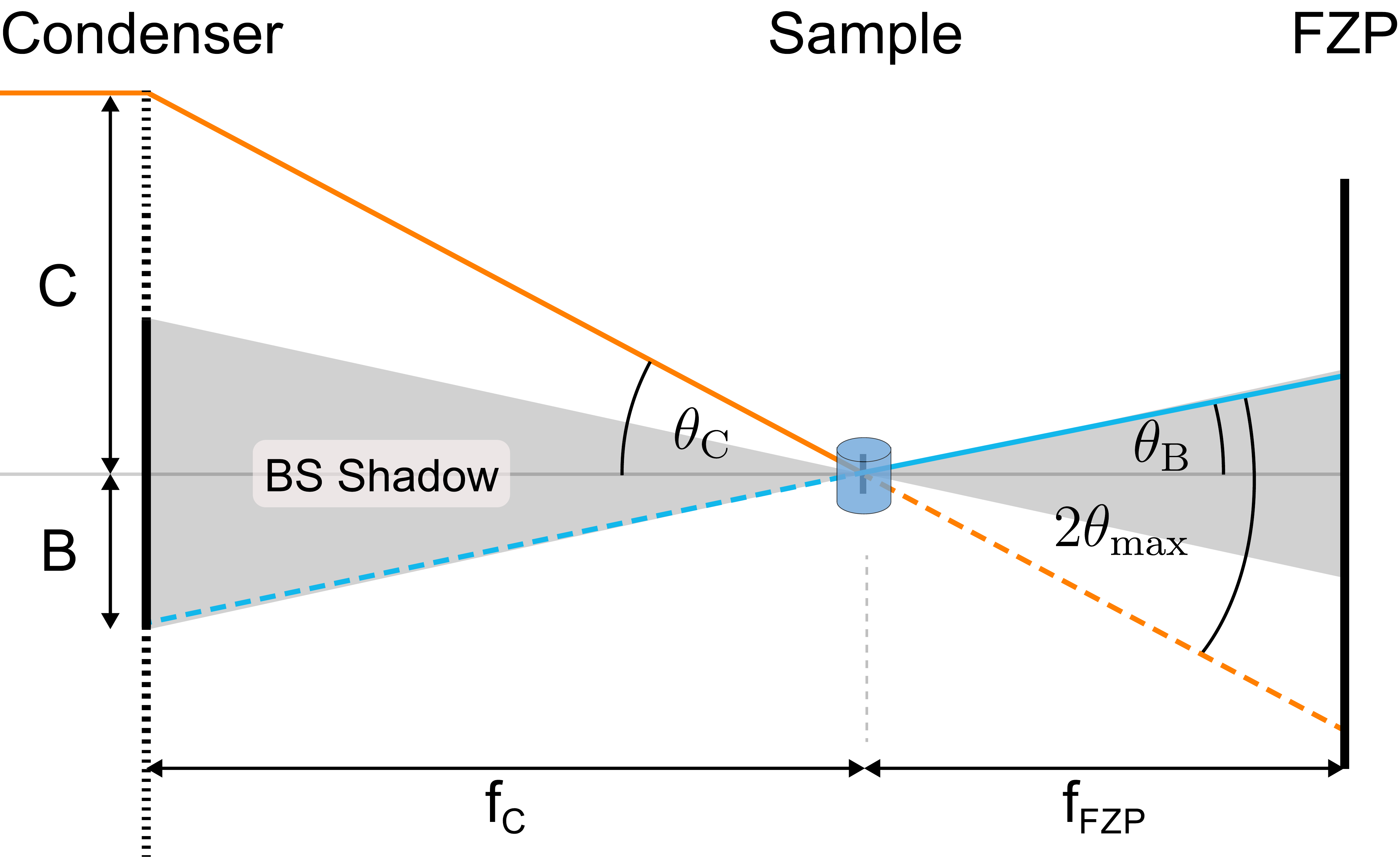}
    \caption{The maximum of the detectable scattering angle is reached, when the outermost beam-let (orange solid line) is scattered towards the opposite side of the beam stop shadow (blue solid line). Scattering with larger angles will be blocked by the apertures and therefore don't contribute to the dark-field signal.}
    \label{fig:scatterin_angle}
\end{figure}

Based on equation \ref{eq:Q}, the scattering vector is limited by the maximal scattering angle accepted by the setup. For the dark-field TXM, the scattering angle $\theta$ becomes largest if the outermost beam-let from the beam-shaping condenser (Figure \ref{fig:scatterin_angle} orange solid line), with the angle $\theta_{\text{C}}$ to the optical axis, is diffracted to the opposite edge of the beamstop shadow (Figure \ref{fig:scatterin_angle} blue solid line), with the angle $\theta_{\text{B}}$. Therefore the maximum detectable scattering angle $\theta_{\text{max}}$, is given by
\begin{align}
    \label{eq:theta}
    2\theta_{\text{max}} = \theta_{\text{C}} + \theta_{\text{B}}.
\end{align}
By using the small angle approximation $\theta_{\text{B}}$ and $\theta_{\text{C}}$ are
\begin{align}
    \label{eq:thetaCthetaB}
    \theta_{\text{C}} \approx \tan{\theta_{\text{C}}} = \frac{C}{f_{\text{C}}}, && \theta_{\text{B}} \approx \tan{\theta_{\text{B}}} = \frac{B}{f_{\text{C}}},
\end{align}
with the focal distance of the condenser $f_{\text{C}}$, the distance of the outermost condenser field to the optical axis $C$, and the radius of the beam stop $B$.
Substituting equation \ref{eq:theta} and \ref{eq:thetaCthetaB} into the equation \ref{eq:Q} gives
\begin{align}
    |\Vec{Q}_{\text{max}}| = \frac{2 \pi}{\lambda f_{\text{C}}} (B + C).
    \label{eq:Q_max}
\end{align}
The focal distance of the beam-shaping condenser also depends on the wavelength, the diameter, and the outermost zone width $\Delta r_{\text{C}}$ of the beamshaping condenser
\begin{align}
    f_{\text{C}} = \frac{2C \Delta r_{\text{C}}}{\lambda}.
    \label{eq:f_C}
\end{align}
Substituting equation \ref{eq:f_C} in equation \ref{eq:Q_max} yields the final expression
\begin{align}
    |\Vec{Q}_{\text{max}}| = \frac{\pi}{\Delta r_{\text{C}}} \left(\frac{B}{C} + 1 \right).
    \label{eq:Q_max_f}
\end{align}

\begin{figure}[htbp]
    \centering
    \includegraphics[width = \textwidth]{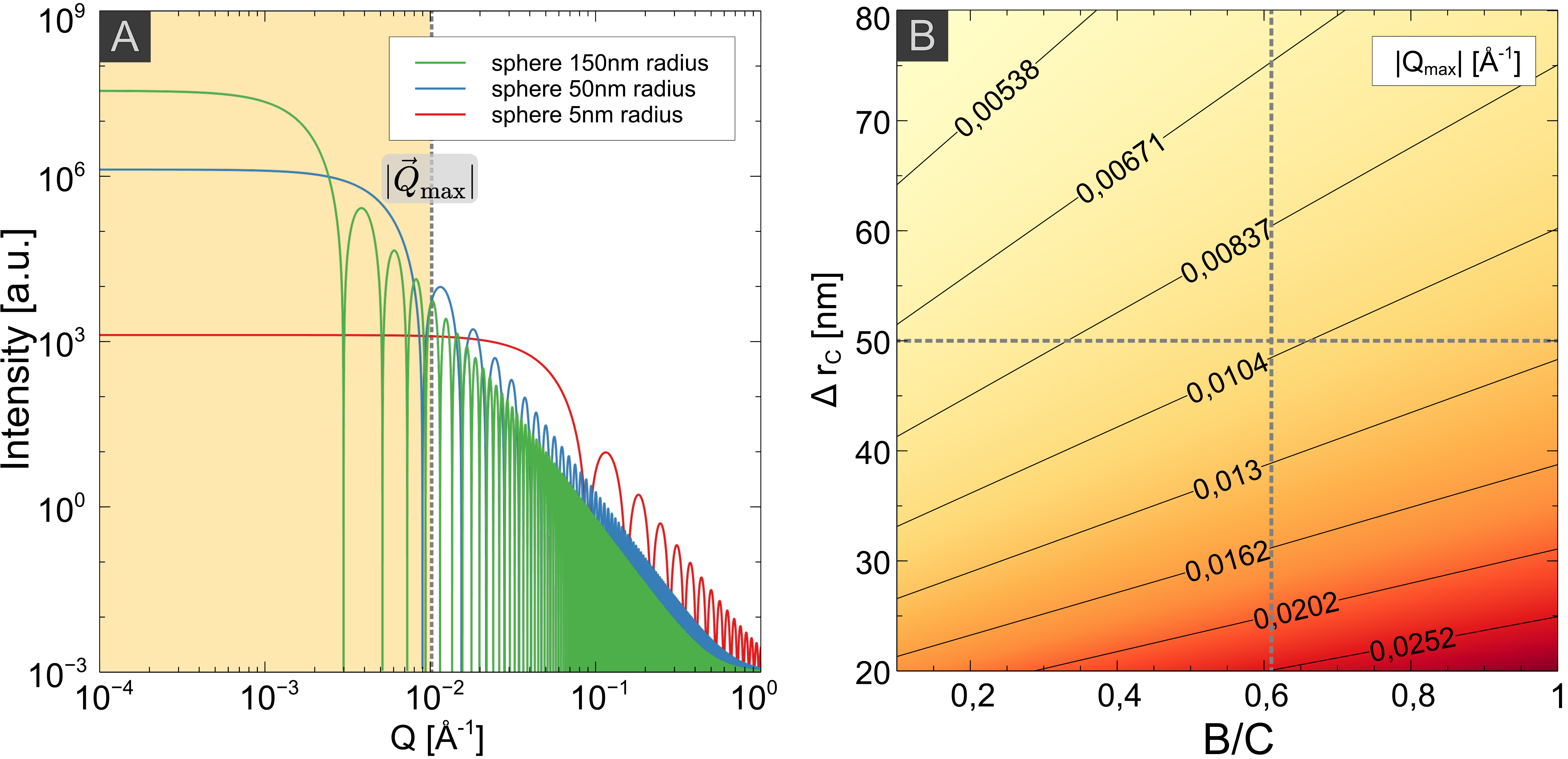}
    \caption{\textbf{A:} The isotropic form factor intensity for spheres with radii of \SI{5}{\nano\meter}, \SI{50}{\nano\meter}, and \SI{150}{\nano\meter}. The vertical grey line marks $|\Vec{Q}_{\text{max}}|$ of the setup. So, only the scattering intensities left of the vertical line (shaded in yellow) contribute to the dark-field image. Smaller structures shift the scattered intensity towards higher q-values but still contribute intensity for small q-ranges.
    \textbf{B:} Contour plot of $|\Vec{Q}_{\text{max}}|$ for the outermost zone width of the beam-shaping condenser $
    \Delta r_{\text{C}}$ over the ratio of the beam stop radius to beam-shaping condenser radius $B/C$. 
    The grey dotted horizontal and vertical lines mark the values of the setup used. }
    \label{fig:I_over_Q}
\end{figure}

In scattering experiments, the smallest resolvable periodic structure is estimated by

\begin{align}
    d = \frac{2\pi}{|\Vec{Q}_{\text{max}}|} = \frac{\lambda f_{\text{C}}}{B + C} = \frac{2C \Delta r_{\text{C}}}{B + C} .
\end{align}
It defines the size of a periodic structure for which the first peak of the resulting scattering intensity is still detectable and can therefore be used to infer the structure of a sample. However, this expression does not necessarily describe the smallest structures detectable with the dark-field TXM. This is because smaller structures shift the intensities to higher q-values in the reciprocal space but still contribute to the scattered intensity in the low q-range of the dark-field setup and therefore contribute to the dark-field image (Figure~\ref{fig:I_over_Q} A). The total scattered intensity passing through the aperture is therefore the limiting factor. It decreases for smaller structures and eventually becomes indistinguishable from the background. 
Increasing $|\Vec{Q}_{\text{max}}|$ still helps to increase the detectability of small structures by allowing more of the intensity, scattered towards high q-values, to pass through to the detector. $|\Vec{Q}_{\text{max}}|$ can be increased by decreasing the outer most zone width of the beam-shaping condenser $\Delta r_{\text{C}}$. Increasing the ratio of $B/C$ will also increase $|\Vec{Q}_{\text{max}}|$, but it reduces the total flux present in the system by a factor equal to its squared inverse. Hence, the size of the beamstop needs to be chosen carefully based on the sample properties. Figure \ref{fig:I_over_Q} B also shows that the presented experimental setup would benefit more from reducing $\Delta r_{\text{C}}$ than increasing $B/C$.
The parameters for the experimental dark-field TXM are presented in table \ref{tab:DFTXM}.

\begin{table}
    \caption{Experimental setup parameters for the dark-field TXM}
    \setlength{\tabcolsep}{3pt}
    \centering
    \begin{tabular}{|c|c|c|c|c|c|c|c|}
        \hline
        effective pixel size & spatial resolution & FoV & $B$ & $C$ & $\Delta r_{\text{C}}$ & $|\Vec{Q}_{\text{max}}|$ & $d$\\
        \hline
        \SI{45.54}{\nano\meter} & \SI{114.42}{\nano\meter} & \SI{93.27}{\micro\meter}& \SI{0.55}{\milli\meter} & \SI{0.9}{\milli\meter} & \SI{50}{\nano\meter} & \SI{0.0102}{\per\angstrom} & \SI{61.72}{\nano\meter}\\
        \hline
    \end{tabular}
    \label{tab:DFTXM}
\end{table}

\subsubsection{Dark-Field Signal}
The dark-field signal $D$ corresponds to the intensity that is scattered by the sample $I_{\text{s}}$, divided by the background scattered intensity $I_{\text{b}}$, which mainly originates from X-rays scattering in air:

 \begin{align}\label{eq:D}
    D = \frac{I_{\text{s}}}{I_{\text{b}}}.
\end{align}

It delivers values larger than one. The dark-field signal of a scattering sample with the thickness $t$ can hence be described similarly to the Beer-Lambert law:

 \begin{align}\label{eq:Deps}
    D = \exp{\left(\int^{t}_{0}\epsilon dt\right)},
\end{align}

with $\epsilon$ being the dark-field coefficient~\cite{Gassert2021, Taphorn2023}. To gain the dark-field coefficient, the dark-field projections can hence be processed and reconstructed very analog to the normal attenuation signal.

\section{Experimental Results}
\begin{figure}[htbp]
    \centering\includegraphics[width = \textwidth]{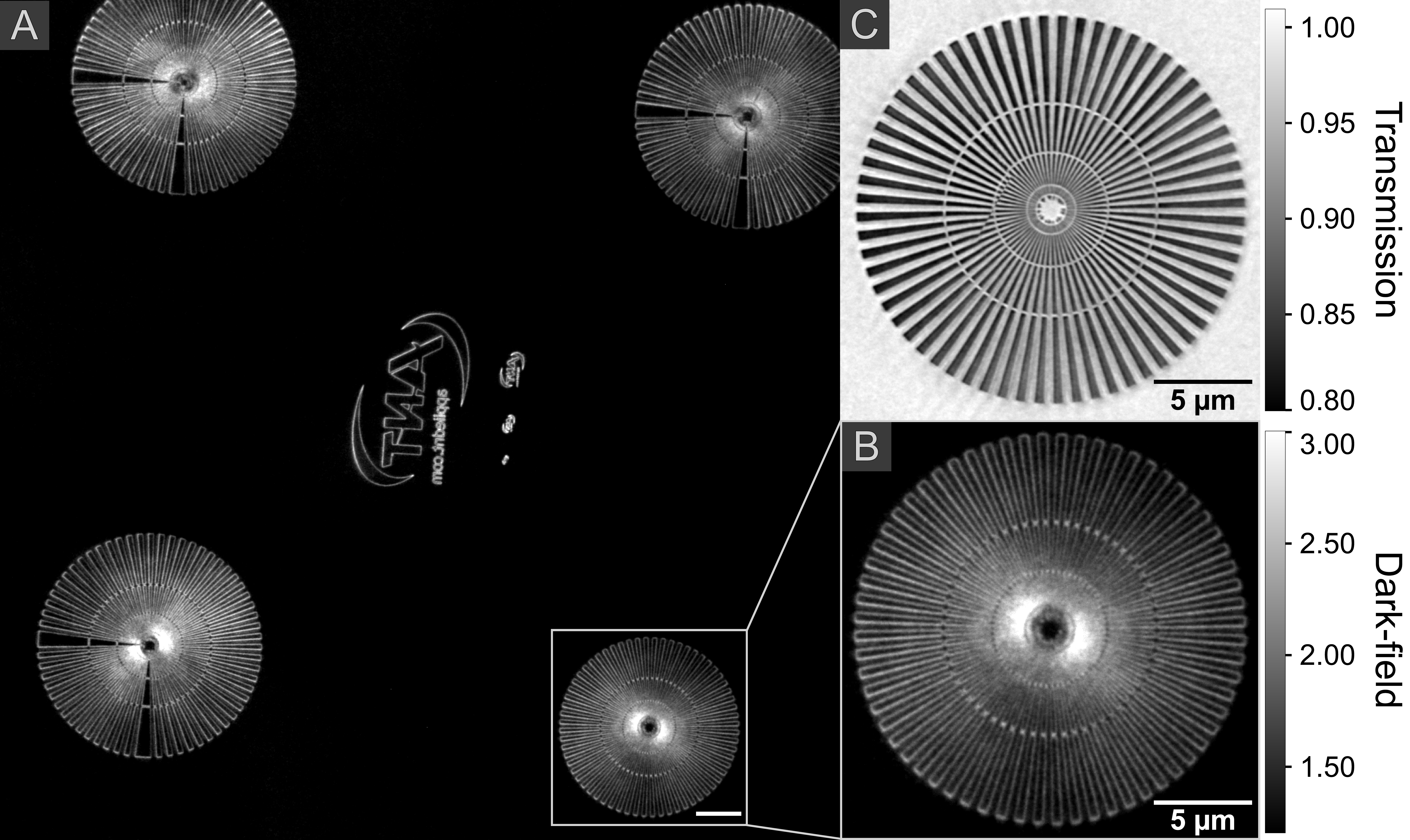}
    \caption{\textbf{A} shows the dark-field of a Siemens star test pattern, with a structure height of \SI{600}{\nano\meter} gold. \textbf{B} shows a close-up of a star, illustrating the frequency dependence of the dark-field signal. \textbf{C} shows the same Siemens star as \textbf{B}, but in transmission after opening the dark-field apertures. Each projection has a total exposure time of \SI{50}{\second}.
    Note: The left part of the Siemens star in \textbf{C} shows a discontinuity which is the result of a camera saving error during the measurement.} 
    \label{fig:siemens_star}
\end{figure}

A Siemens star test pattern, made out of gold, with a structure height of \SI{600}{\nano\meter} and structure sizes down to below \SI{50}{\nano\meter} is measured. By closing the apertures in the BFP, the imaging modality of the setup is switched from transmission (Figure \ref{fig:siemens_star} C) to dark-field (Figure \ref{fig:siemens_star} B). The edges of the test pattern lead to a strong change in the electron density due to the transition from gold to air and, therefore, scatter X-rays. These edges appear bright in the dark-field projection as shown in Figure \ref{fig:siemens_star} A and B. The structures are fully resolved for large radii, and all edges of a star spike light up. Moving towards the center of the Siemens star, towards smaller structures, the dark-field signal increases due to a higher density of scattering structures. The contrast between the structures decreases towards the center due to the resolution limit of the setup, showing a frequency dependence of the signal. The dark-field also visualizes scattered intensity from sub-resolution structures, whereas this information is otherwise lost in attenuation and Zernike phase contrast.

\begin{figure}[htbp]
    \centering\includegraphics[width = 0.75\textwidth]{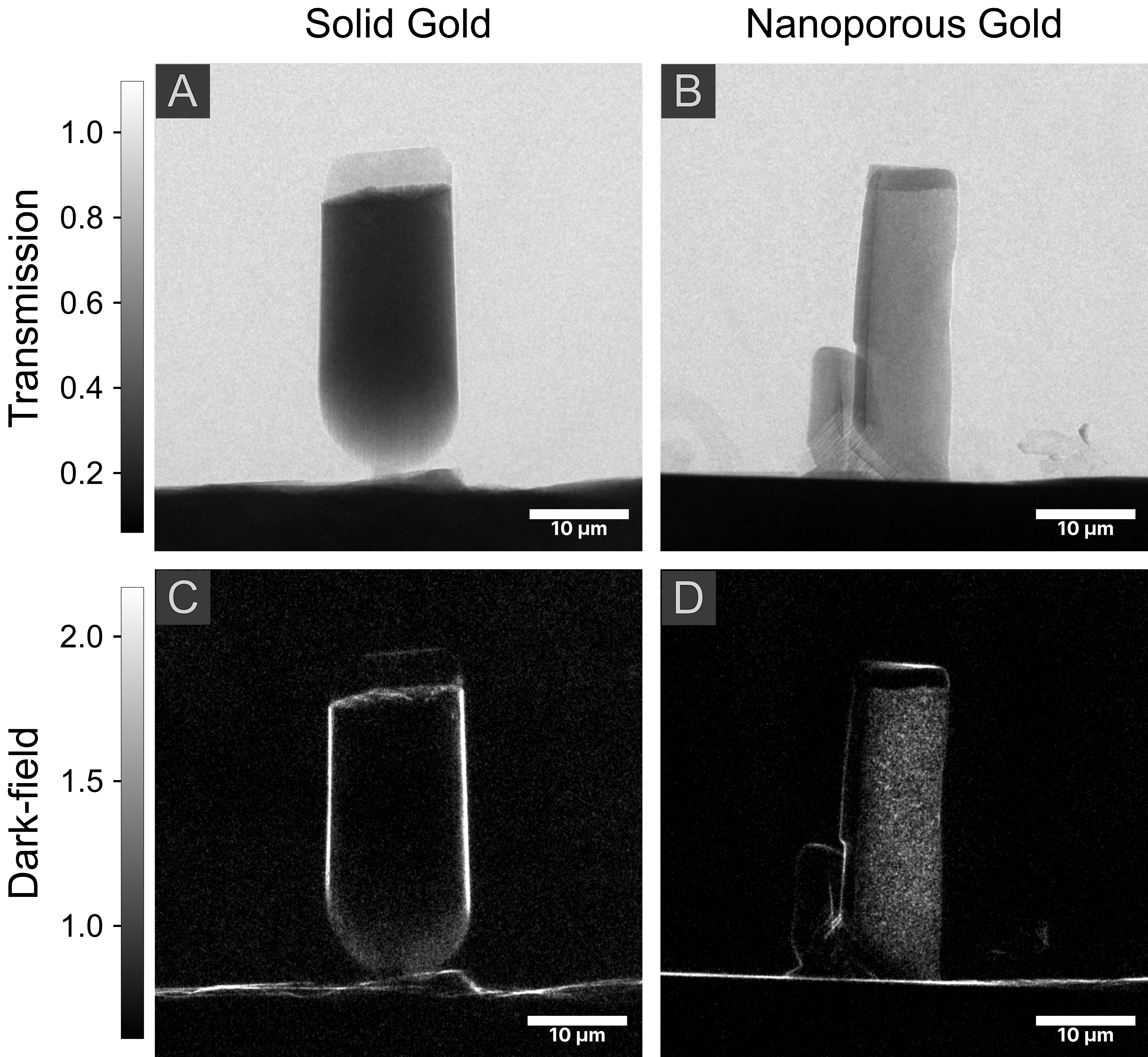}
    \caption{Two gold pillars measured in the bright field (\textbf{A} and \textbf{B}) and in the dark-field (\textbf{C} and \textbf{D}). The pillar in \textbf{A} and \textbf{C} consists of solid gold, while the pillar in \textbf{B} and \textbf{D} contains a nanometer hierarchical inner structure. In the transmission projections, both pillars look similar, whereas the dark-field projection reveals a strong difference in the inner structure of the two pillars. Each projection was taken with \SI{1}{\second} exposure time.} 
    \label{fig:gold_pillar}
\end{figure}

The dark-field signal can also help to improve detectability, e.g. of cracks and pores, in certain applications.
Figure \ref{fig:gold_pillar} shows two gold pillars. One pillar is made out of solid gold, while the other contains a nanometer hierarchical inner structure in the range of \SI{120}{\nano\meter} and below, which is at the resolution limit of the TXM setup used. Therefore, the transmission projection of both pillars looks homogeneous within the sample (Figure \ref{fig:gold_pillar} A, B), and based on the projection alone, it is hard to make a statement about its inner structure. In contrast, the dark-field projection immediately reveals a substantial difference between the two pillars since the nanoporous structure strongly scatters the X-rays, while the solid pillar shows only a scattering signal at the edges (Figure \ref{fig:gold_pillar} C, D). In the dark-field, the separation of the nanoporous gold structure from the solid structure next to it is also visible. In both dark-field projections (Figure \ref{fig:gold_pillar} C, D), the background drops below a value of one. This phenomenon can be attributed to the fact that the samples are mounted on a solid metal base, which blocks a significant part of the air-scattered X-rays. For the reference image, the sample, including the metal base, is moved outside the beam. Therefore, the background scattered intensity $I_{\text{b}}$ in equation \ref{eq:D} is larger than $I_{\text{s}}$ for pixels outside the sample, resulting in values below one.

\begin{figure}[htbp]
    \centering\includegraphics[width = \textwidth]{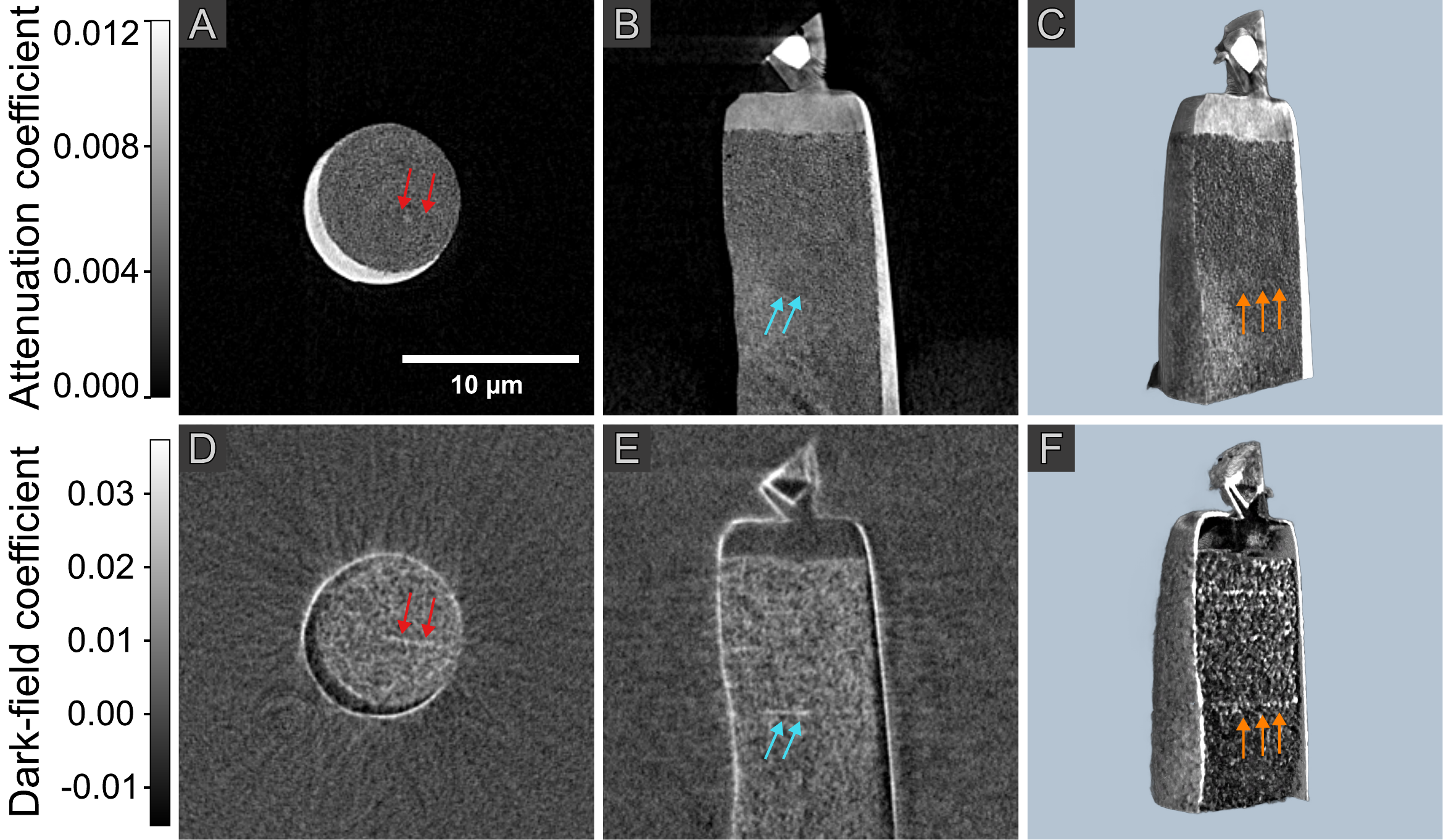}
    \caption{Reconstructed volume of a nanoporous gold pillar. \textbf{A}, the axial and \textbf{B}, the coronal view show the local distribution of the attenuation coefficient. \textbf{D} and \textbf{E} show the local distribution of the dark-field coefficient. \textbf{C} and \textbf{F} are the corresponding volume renderings. The arrows indicate the grain boundaries, which are visible in the scattering but not in the attenuation coefficient. Each tomogram was acquired in less than \SI{15}{\minute}.}
    \label{fig:dark_field_tomo}
\end{figure}

The use of a beam-shaping condenser in combination with the long propagation distance at the P05 beamline results in a high efficiency of the TXM setup. The beam-shaping condenser allows a large area of the initial beam to be used to illuminate the sample. The long propagation distance allows purely geometric magnification, eliminating the need for an additional, less efficient light optical microscope.
This allows short exposure times and fast tomography scans, which also applies to the dark-field TXM. Figure \ref{fig:dark_field_tomo} shows the reconstructed volume of the nanoporous gold pillar shown above, acquired in less than \SI{15}{\minute}, providing information about the local distribution of the attenuation (A, B, C) and dark-field coefficients (D, E, F). The volume of the dark-field coefficient reveals substructures indicated by the arrows, which are not visible in the attenuation coefficient. These structures are likely to be grain boundaries. Both tomograms were performed by a \SI{180}{\degree} scan over 1588 projections using a high precision rotation stage~\cite{Flenner2020}. Each projection was exposed for \SI{0.5}{\second}, resulting in a total exposure time per tomogram of \SI{13.24}{\minute} excluding references.

\section{Discussion}
The use of motorized apertures in the BFP of the FZP allows rapid switching between the transmission and dark-field modality without refocusing, minimizing downtime and maximizing data acquisition efficiency. The use of a beam-shaping condenser in combination with purely geometric magnification allows short exposure times of only \SI{0.5}{\second} for dark-field projections. This enables tomographic scans to be acquired in less than \SI{15}{\minute} (Figure \ref{fig:dark_field_tomo}). The high throughput offers potential for applications where rapid imaging is critical, such as real-time studies of material deformation or observation of rapidly evolving processes like the crystallization of materials~\cite{John2024}. Furthermore, motorized apertures can be positioned precisely, making the method easy to implement, reducing experimental complexity, and improving reproducibility. Instead of an aperture, one could also use absorption rings to block the bright field, similar to the Zernike phase contrast method \cite{Zernike1942, Schmahl1994, Neuhausler2003, Stampanoni2010, Flenner2023} but with higher gold structures. This would allow more of the scattered light to be used, increasing the measured intensity. However, using absorption rings instead of phase rings is difficult because the differential phase shifts in the sample lead to a shift of the focal spots in the BFP. If this shift is greater than the width of the absorption ring, parts of the bright field can pass through to the detector and outshine the dark-field signal. The width of the absorption ring must, therefore, be carefully chosen based on the sample properties. This approach is also much more sensitive to beam instabilities, as small positional variations can be enough to cause the bright field and absorption rings to be out of alignment.

From scattering experiments, a mathematical expression is derived that describes the response of the setup to the scattering vector and allows size estimation of the smallest structures that can be detected. This value, however, does not describe the hard limit of the setup. In SAXS, the scattered intensities are distributed continuously in the reciprocal space. Even though the first peak of the form factor lies outside the detectable range of the setup, the dark-field projection can still contain information about the sample structure as shown in Figure \ref{fig:I_over_Q} A. The dark-field signal is, therefore, mainly limited by the flux of the imaging system and the efficiency of the detector. Here, the experiments are performed with a charge-integrating detector. Instead, one could also use photon-counting detectors with higher efficiency~\cite{Flenner2023}. The larger pixel size of the photon-counting detectors normally means a reduction in resolution and, with that, loss of information. However, the dark-field signal is independent of the pixel size. Even though the spatial resolution of the projection will be lower, smaller structures can still be detected inside the area of a pixel by the dark-field signal. Hence, the use of photon-counting detectors helps to increase the detectability of small structures even though they are not directly resolvable.

The applicability of the presented approach has been demonstrated by measuring a hierarchical nanoporous gold pillar, whose intrinsic structure can be directly visualized in projections using the dark-field signal (Figure \ref{fig:gold_pillar}). We also showed that the dark-field coefficient allows the detection of sample structures that are not visible in the attenuation coefficient (Figure \ref{fig:dark_field_tomo}). However, the projection of the Siemens stars also shows an angular dependence of the dark-field signal (Figure \ref{fig:siemens_star}). In the diagonal directions, the dark-field signal is brighter. There are two possible reasons for this: First, it could be due to the circular illumination pattern in the BFP coming from the beam-shaping condenser, but using a rectangular aperture (see Figure \ref{fig:setup} B). Therefore, more scattered light can pass through the apertures in the diagonal directions. A circular aperture instead, would lead to a uniform occluding of the focal spots in the BFP, and the distance from the outermost focal spots to the opposite edge of the circular aperture would be independent of the directionality.
The second reason could be inconsistencies in the optical system. Figure \ref{fig:siemens_star} C shows the transmission projection of one of the Siemens stars, which appears slightly blurred in the same direction where the dark-field shows a reduced intensity, which would support the latter.

A scanning approach to utilize the scattered intensity in X-ray microscopy is provided by H. Simons \textit{et al.} \cite{Simons2015} and C. Yildirim \textit{et al.} \cite{Yildirim2020}. They use off-axis optics to focus the scattered light onto a detector. The setup is used for three-dimensional imaging of orientations and stresses within sample volumes. It also utilizes only the scattered intensity to create an image of the sample. However, the off-axis arrangement of the optics allows only a limited amount of scattered intensity to be used, since only the intensity scattered towards the optics is detected. Intensity scattered in other directions is lost. This makes the setup a scanning method that relies on the Bragg condition within the sample. Although this allows the mapping of the 3D orientations by rotating the sample around two rotation axis, it reduces the effectiveness for measurements where only the 3D visualization of the scattering sources is of interest. Therefore, their method can be clearly distinguished from the full-field dark-field TXM method presented here.

\section{Conclusion}
While X-ray dark-field imaging is increasingly used in laboratory and synchrotron-based micro-CT systems, there is a lack of methods for nanometer-scale applications. 
Here, a full-field dark-field TXM setup is proposed to visualize the scattering structures of a sample. The technique is able to extend the detection of sample features below the setup's imaging resolution, independent of the pixel size, while maintaining short exposure times. The dark-field signal is sensitive to small inner structures with strong changes in the electron density, like cracks, bubbles, or material boundaries. The additional information is especially helpful in the field of material science, where the inner structures are crucial to gain a deeper understanding. 
This is the first time, that full-field dark-field TXM has been demonstrated using apertures in the BFP. Due to the simple implementation, the proposed method can easily be used to extend existing setups by the dark-field signal, which could make dark-field imaging broadly available for nanometer X-ray imaging.

\begin{backmatter}

\bmsection{Funding} We acknowledge financial support by Deutsche Forschungsgemeinschaft (DFG), ResearchTraining Group, under Grant GRK 2274.
The authors gratefully acknowledge financial support from the Deutsche Forschungsgemeinschaft (DFG) (project No. 192346071, SFB 986, project Z2).

\bmsection{Disclosures} The authors declare no conflicts of interest.

\bmsection{Data availability} Data underlying the results presented in this paper are not publicly available at this time but may be obtained from the authors upon reasonable request.

\bmsection{Acknowledgments} The authors thank Martin Ritter for providing the nanoporous sample and Martin Dierolf, Malte Storm, Dominik John, and Kirsten Taphorn for fruitful discussions.

\end{backmatter}


\bibliography{sample}






\end{document}